\documentclass{njcarticle}
\usepackage{txfonts}
\titlehead{ Streaming Maximum-Minimum Filter}
\authorhead{Daniel Lemire}
 \title{ \MakeUppercase{Streaming Maximum-Minimum Filter Using No More than Three Comparisons per Element}}
\author{Daniel Lemire\\ 
University of Quebec at Montreal (UQAM), UER ST \\
100 Sherbrooke West, Montreal  (Quebec), H2X 3P2 Canada\\
{\small \tt lemire@acm.org}
}
 
\usepackage{listings} 
\usepackage{amssymb}
\usepackage{algorithm}
\usepackage[noend]{algorithmic} 
\usepackage{subfigure}
\usepackage{natbib}
\newcommand{\komment}[1]{}
\usepackage{ifpdf}
\usepackage{color}
\ifpdf
    \providecommand{\myfig}[1]{#1.pdf}
   \usepackage[pdftex]{graphicx}
\else
   
\providecommand{\myfig}[1]{#1.eps}
   \usepackage{graphicx}
\fi

\usenjcepsmacros
\njccitationstyle

\twolevelnumbering{equation,figure}
\draft

\begin{document}

\maketitle
%
%
%


\begin{abstract}
The running maximum-minimum (\textsc{max-min}) filter computes the 
maxima and minima over running windows of size $w$. 
This filter has numerous applications in signal processing and
time series analysis. We present an easy-to-implement online algorithm
requiring no more than 3~comparisons per element, in the worst case. 
Comparatively, no algorithm is known to compute the running
maximum (or minimum) filter in  1.5~comparisons per element, 
in the worst case. Our algorithm has reduced latency and memory usage. 
\end{abstract}
\begin{subject}F.2.1 Numerical Algorithms and Problems\end{subject}

\maketitle


\section{Introduction}

The maximum and the minimum are the simplest form of order statistics.  Computing either
the global maximum or the global minimum  of an array of $n$ elements 
requires $n-1$~comparisons, or slightly less than
one~comparison per element. However,
to compute simultaneously the maximum and the minimum, 
only $3 \lceil n/2 \rceil -2$~comparisons are required in the worst case~\bracketcite{corman2001ia}, or slightly less 
than 1.5~comparisons per element. 

\begin{figure}
\begin{center}
\includegraphics[width=0.4\columnwidth,angle=270]{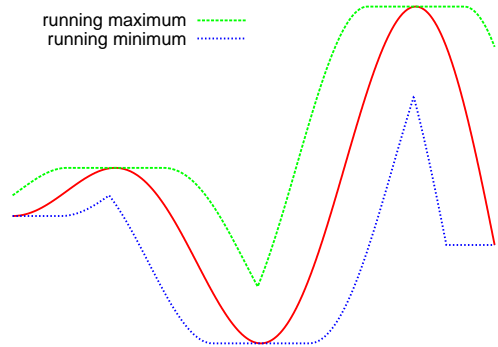}
\end{center}
\caption{
\label{maxminexample}Example of a running \textsc{max-min} filter.
}
\end{figure}

A related problem is the computation of the running maximum-minimum 
(\textsc{max-min}) filter: given an array $a_1, \ldots, a_n$, find the maximum and the minimum over all 
windows of size $w$, that is $\max/\min_{i\in [j,j+w)} a_i$ for all $j$ (see \figref{maxminexample}).
The running maximum (\textsc{max}) and minimum (\textsc{min}) filters are defined similarly.
The \textsc{max-min} filter problem is harder than the  \textsc{global max-min} problem, but a tight bound on the number of comparisons required in the worst
case remains an open problem.

Running maximum-minimum (\textsc{max-min}) filters are used in signal processing and
pattern recognition. As an example, \cite{keogh2005eid} use a precomputed
\textsc{max-min} filter to approximate the time warping distance between two time series.
Time series applications range from  music retrieval~\bracketcite{zhu2003wie}
to network security~\bracketcite{sun2004dal}.
The unidimensional \textsc{max-min} filter can be applied to
images and other bidimensional data by first applying the
 unidimensional on rows and then on columns. Image processing applications 
include cancer diagnosis~\bracketcite{He2005}, character~\bracketcite{ye2001smb} and handwriting~\bracketcite{Ye2001}
recognition, and boundary feature comparison~\bracketcite{Taycher2004}.


 We define the \textit{stream latency} of a filter as the maximum number of data points required
after the window has passed. For example, an algorithm requiring that the whole
data set be available before the running filter can be computed has a 
high stream latency. In effect, the stream latency is a measure of an algorithm
on the batch/online scale. 
We quantify the speed of an algorithm by the number of comparisons between values,
either $a<b$
or $b<a$, where values are typically floating-point numbers.
 
We present  the first algorithm to compute the combined \textsc{max-min} filter in no more than
3~comparisons per element, in the worst case. Indeed, we are able to save some comparisons
by not treating the \textsc{max-min} filter as the aggregate of the \textsc{max} and
\textsc{min} filters: if $x$ is strictly larger than $k$ other numbers,
then there is no need to check whether $x$ is smaller than any of these numbers.
 Additionally, it is the
first algorithm to require a constant number of comparisons per element 
without any stream latency and it uses less memory than competitive alternatives. 
Further, our algorithm requires
no more than 2~comparisons per element when the input data is monotonic (either non-increasing or non-decreasing). 
We provide experimental evidence that our algorithm is competitive and can
be substantially faster (by a factor of~2) when the input data is piecewise
monotonic.
A maybe surprising result is that our algorithm is arguably  simpler to implement
than the recently proposed algorithms such as \cite{628856} or \cite{1059644}. 
Finally, we prove that at least 2~comparisons per element are required to compute the \textsc{max-min}
filter when no stream latency is allowed.

%
%
%
%
%
%
%
%
%
%
%

\section{Related Work}

\cite{pitas1989far} presented the \textsc{max} filter 
algorithm \textsc{maxline} requiring 
$O(\log w)$ comparisons per element in the worst case and an average-case performance 
over independent and identically distributed (i.i.d.) noise data of slightly more
than 3~comparisons per element. 
\cite{douglas1996rmm}  presented a
better alternative: the \textsc{max} filter algorithm \textsc{maxlist} was shown to average 3~comparisons per element for i.i.d. input signals and \cite{myers1997aim} 
presented 
an asynchronous implementation.

\begin{figure}
\begin{center}
\includegraphics[width=0.6\columnwidth,angle=270]{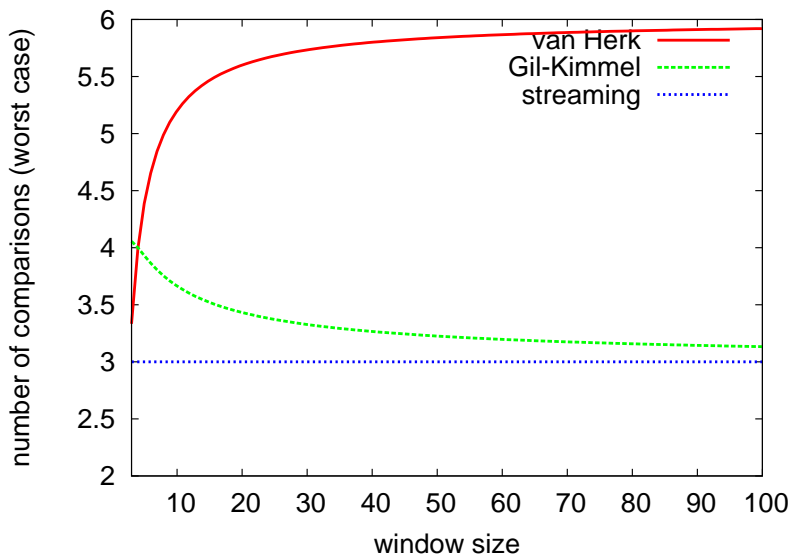}
\end{center}
\caption{
\label{theory}Worst-case number of comparisons per element with  the \textsc{van Herk-Gil-Werman} (van Herk) algorithm, the \textsc{Gil-Kimmel} algorithm, and our new streaming algorithm (less is better). 
}
\end{figure}

More recently, \cite{139337} and \cite{628474} presented an algorithm  requiring $6-8/w$~comparisons
per element,  
in the worst case. 
The algorithm
is based on the batch computation of cumulative maxima and minima over overlapping
blocks
of $2w$ elements. For each filter (\textsc{max} and \textsc{min}), it uses
a memory buffer of $2w+O(1)$~elements.  We will
refer to this algorithm as the \textsc{van Herk-Gil-Werman} algorithm.
\cite{628856} proposed an improved version (\textsc{Gil-Kimmel}) which lowered the number of comparisons
per element to slightly more than 3~comparisons per element, but at the cost
of some added memory usage and implementation complexity
(see \tableref{bigtable} and \figref{theory} for summary). 
For i.i.d. noise data,  Gil and Kimmel presented a variant of
the algorithm requiring $\approx 2+(2+\ln 2 /2)\log w/w$~comparisons per element
(amortized), but with the same worst case complexity. Monotonic data
is a worst case input for the \textsc{Gil-Kimmel} variant.

\begin{table}
\centering
\caption{\label{bigtable}Worst-case number of comparisons and stream latency for competitive \textsc{max-min} filter algorithms. Stream latency and memory usage (buffer) are
 given in number of elements.}
\begin{small}\begin{tabular}{p{3cm}p{3cm}cc}
\hline
algorithm   &  comparisons per element (worst case) & stream latency & buffer\\ 
\hline \hline
\centering naive  & \centering$2w-2$ &  \textbf{0} & \textbf{$O(1)$} \\ \hline
\centering\cite{139337}, \cite{628474}  & \centering$6-8/w$ & $w$ & $4w+O(1)$ \\ \hline
\centering\cite{628856}  & \centering $3 + 2 \log w /w$ +$O(1/w)$ &$w$ & $6w+O(1)$ \\ \hline
\centering\textbf{New algorithm} & \centering\textbf{3}& \textbf{0} & $2w+O(1)$\\
\hline 
\end{tabular}             \end{small}
\end{table}

%
%
%
%
%
%

\cite{1059644} proposed a fast algorithm based on anchors. They do not
improve on the number of comparisons per element. 
For window sizes ranging from 10 to 30 and data values ranging from 0 to~255,
their implementation has a running time lower than 
their \textsc{van Herk-Gil-Werman} implementation by as much as 30\%. Their \textsc{Gil-Kimmel} implementation outperforms
their \textsc{van Herk-Gil-Werman} implementation by as much as 15\% for window sizes 
larger than 15, but is outperformed similarly
for smaller window sizes, and both are comparable for a window size equals to 15.
The Droogenbroeck-Buckley \textsc{min} filter pseudocode  alone requires a full page compared to 
a few lines for  \textsc{van Herk-Gil-Werman} algorithm. Their experiments did not consider
window sizes beyond $w=30$ nor arbitrary floating point data values.



%
%

\komment{\section{The \textsc{van Herk-Gil-Werman} Algorithm and Variants}
The \textsc{van Herk-Gil-Werman} algorithm can compute either the
running $\max$ or the $\min$ filter using $3-4/w$~comparisons
per element. By combining both filters, the $\textsc{max-min}$ filter
is computed using  $6-8/w$~comparisons
per element. 

The maximum over a window
can be decomposed into a prefix and a suffix maximum around a multiple of the
window width ($w$):
$\max_{i\in [j,k)} a_i = \max \{ \max_{i\in [j,\lfloor k/w\rfloor w)} a_i, \max_{i\in [\lfloor k/w\rfloor w+1,k)} a_i \}.$
The prefixes and suffixes contained in any  block ($[1,w],[w,2w],\ldots$) can be computed using
 $2w-2$~comparisons.  
Moreover, there are $w$~windows 
starting in each block  ($[1,w),[w,2w),\ldots$), and since two of them per block
are already a suffix or a prefix (for example $[1,w]$ and $[2,w+1]$),
we require $w-2$~comparisons per block to compare the prefixes and the suffixes. In total, we require $3w-4$~comparisons per block or $3-4/w$~comparisons
per element.

\cite{628856} improved this algorithm in two ways:
\begin{itemize}
 \item The windows ranging from $[1,w]$ to $[w,2w-1]$ are computed as the maximum of
a prefix (in $[1,w+1]$) and a suffix (in $[w+1,2w+2]$), but the values
of the prefixes are monotonically decreasing and the values of the suffixes
are monotonically increasing. While initially, the prefixes are larger,
at one point, the suffixes will be larger.
 So, instead of spending $w-2$~comparisons by taking the
maximum of each suffix with each prefix, we can use a binary search which costs
only  $\lceil \log ( w- 1 )\rceil$~comparisons to find  when the suffixes 
become larger than the prefixes. This variant alone means that the
number of comparisons per element is reduced to  $2-2/w+\log ( w-1)/w$.
\item For a given block, we can compute the prefixes and suffixes simultaneously
using less than $2w-2$~comparisons. If we knew
 where the global
maximum was in the block, we could simply compute the prefix and suffix
 maxima up to that point, since cumulative maxima become constant after
the global maximum is passed. We do not know where the global
maximum is, but we know that it 
is in either the first or second half of the block. 
Hence, we only compute the first $q=\lceil (w+1)/2\rceil$ prefix maxima
(for example $[1,1]$, $[1,2]$, \ldots, $[1,q]$)
and  the first $w- \lceil (w+1)/2\rceil$ suffix maxima (for example
$[w+1,w+1]$, $[w,w+1]$, \ldots, $[q+1,w+1$]).
By comparing the last prefix and the last suffix maxima to determine
in which half the global maximum is. If the global maximum is in the first
half, the prefixes do not need to be computed in the second half, and vice versa.
This reduces the number of comparisons to  $1.5 w- w\bmod{2}/2$ 
from $2w-2$.
Combining this variant with the previous one, the number of comparisons
per element
is  $1.5+\log  w/w+O(1/w)$.
\end{itemize}
Therefore, \textsc{Gil-Kimmel} computes the \textsc{max-min} filter 
 using no more than $3+2\log w / w$~comparisons.
As the window size grows large ($w\rightarrow \infty$), the number
of comparisons goes down to 3.

These algorithms are particularly competitive when the window size $w$ is large,
but then they introduce a significant latency: between the moment
where a window has passed, and the moment the algorithm outputs its maximum and minimum,
we must wait for up to $w$ additional data points. Their memory usage is also larger.}
  

%
%

%
%
\section{Lower Bounds on the Number of Comparisons}

\cite{628856} showed 
that the \textsc{prefix max-min} ($\max/\min_{i\leq j} a_i$ for all $j$)
requires at least $\log 3\approx 1.58$~comparisons per element,  while they conjectured
that at least 2~comparisons are required. We prove that their result
applies directly to the \textsc{min-max} filter problem and show that
2~comparisons per element are required when no latency is allowed.

%
%

\begin{theorem}\label{bigtheorem}
In the limit where the size of the array becomes infinite,
 the \textsc{min-max} filter problem  requires at least 2~comparisons per element when no stream latency 
is allowed, and  $\log 3$~comparisons per element otherwise.
\end{theorem}
\begin{proof}
Let array values be distinct real numbers.
When no stream latency is allowed, we must return the maximum and minimum of
window $(i-w,i]$ using only the data values and comparisons in $[1,i]$. An adversary can choose
the  array value $a_i$ so that $a_i$ must be compared at least twice
with preceding values: it takes two comparisons with $a_i$ 
 to determine that it is neither a maximum nor a minimum
($a_i \in (\min_{j\in (i-w,i]} a_j, \max_{j\in (i-w,i]} a_j)$). Hence, at least $2(n-w)$~comparisons are
required, but because 
$2(n-w)/n\rightarrow 2$ as $n\rightarrow \infty$, two comparisons
per element are required in the worst case.

Next we assume stream latency is allowed.
Browsing the array from left to right, each new data point $a_i$ for $i\in [w,n]$ 
can be
either a new maximum
($a_i = \max_{j\in (i-w,i]} a_j$), a new minimum 
($a_i = \min_{j\in (i-w,i]} a_j$), or  neither a
new maximum or a new minimum ($a_i \in (\min_{j\in (i-w,i]} a_j, \max_{j\in (i-w,i]} a_j)$).
For any ternary sequence such as MAX-MAX-MIN-NOMAXMIN-MIN-MAX-\ldots, we can generate
a corresponding array.
This means that a \textsc{min-max} filter needs to distinguish
between more than $3^{n-w}$~different partial orders 
over the values in the array $a$.
In other words, the binary decision tree must have more than $3^{n-w}$~leaves.
Any binary tree having $l$~leaves has height at least $\lceil \log l \rceil$.
Hence, our binary tree must have height at least $\lceil \log  3^{n-w}  \rceil\geq (n-w) \log 3$,
proving that $(1-w/n)\log 3 \rightarrow \log 3$~comparisons per element are required when $n$ is large. \end{proof}

By the next
proposition, we show that the general lower bound of 2~comparisons per element is tight.

\begin{proposition}\label{smallprop}
There exists an algorithm to compute the \textsc{min-max} filter in no more than 2~comparisons per element when the window size is 3 ($w=3$), with  no stream latency.
\end{proposition}
\begin{proof}
 Suppose we know the location of the maximum and minimum of the window $[i-3,i-1]$. Then we know the maximum
and minimum of $\{a_{i-2},a_{i-1}\}$. Hence, to compute the maximum and minimum of $\{a_{i-2},a_{i-1},a_i\}$, it suffices
to determine whether $a_{i-1}>a_i$ and whether $a_{i-2}>a_i$.
\end{proof}

\section{The Novel Streaming Algorithm}


%
%
%

\begin{figure}
\begin{center}
          \resizebox{0.8\columnwidth}{!} {
             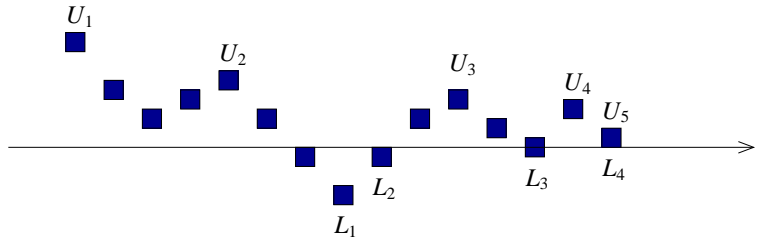
           }
\end{center}
\caption{
\label{monotonicwedge}Example of a monotonic wedge: data points run from left to right. }
\end{figure}

To compute  a running \textsc{max-min} filter, it is sufficient  to maintain
a \textit{monotonic wedge} (see \figref{monotonicwedge}).
Given an array $a=a_1,\ldots,a_n$, a monotonic wedge is made of two lists $U$, $L$ where
 $U_1$ and $L_1$ are the locations of global maximum and minimum,
$U_2$ and $L_2$ are the locations of the global maximum and minimum in $(U_1,\infty)$
and $(L_1,\infty)$, and so on. Formally, $U$ and $L$ satisfy
 $\max_{i> U_{j-1}}a_i=a_{U_j}$ and $\min_{i> L_{j-1}}a_i=a_{L_j}$ for $j=1,2, \ldots$
where, by convention, $U_{0}=L_{0}=-\infty$. If all values of $a$ are distinct,
then the monotonic wedge $U,L$ is unique. The location of the last data point $n$ in $a$, is the last value stored in both $U$ and $L$ (see $U_5$ and $L_4$ in \figref{monotonicwedge}). A monotonic wedge has the property
that it keeps the location of the current (global) maximum ($U_1$) and minimum ($L_1$) while it can be easily updated
as we remove data points from the left or append them from the right:
\begin{itemize}
 \item to
compute a monotonic wedge of $a_2,a_3,\ldots,a_n$ given a monotonic wedge $U,L$ for $a_1,a_2,\ldots,a_n$, it suffices to remove (pop)  $U_1$ from $U$ if $U_1=1$ or $L_1$ from $L$ if $L_1=1$;
\item similarly, to compute the monotonic wedge
of $a_1,a_2,\ldots,a_n,a_{n+1}$, if $a_{n+1}>a_n$, it suffices to remove the last locations stored in $U$ until $a_{\textrm{last}(U)}\geq a_{n+1}$ or else, to remove the last locations stored in $L$ until $a_{\textrm{last}(L)}\leq a_{n+1}$, and then to append the location $n+1$ to both $U$ and $L$.
\end{itemize}

\figref{mainalgo1} provides an example of how the monotonic wedge for window $[i-w,i-1]$
is updated into a wedge for $[i-w+1,i]$. In Step A, we begin with a monotonic 
wedge for $[i-w,i-1]$. In Step B, we add value $a_i$ to the interval. This new value is compared
against the last value $a_{i-1}$ and since $a_i>a_{U_5}$, we remove the index $U_5$ from  
$U$. Similarly, because $a_i>a_{U_4}$, we also remove $U_4$. In Step C, the index
$i$ is appended to both $U$ and $L$ and we have a new (extended)
 monotonic wedge.
Then, we would further remove $L_1$, consider the next
value forward, and so on.

Algorithm~\ref{algo:streaming} and Proposition~\ref{mainprop} show that a monotonic wedge
can be used to compute the \textsc{max-min} filter efficiently and with few lines of code.

\begin{algorithm}
\begin{small}
 \begin{algorithmic}[1]
\STATE \textbf{INPUT:} an array $a$ indexed from $1$ to $n$
\STATE \textbf{INPUT:} window width $w>2$
\STATE $U$, $L$ $\leftarrow$ empty double-ended queues, we append to ``back'' 
\STATE append $1$ to $U$ and $L$ 
\FOR{ $i$ in $\{2,\ldots,n\}$}\label{alg:mainloop}
\IF{$i\geq w+1$}
\STATE \textbf{OUTPUT:} $a_{\textrm{front}(U)}$ as maximum of range $[i-w,i)$ 
\STATE \textbf{OUTPUT:} $a_{\textrm{front}(L)}$ as minimum of range $[i-w,i)$ 
\ENDIF
\IF{$a_i > a_{i-1}$}\label{alg:firstcompare}
\STATE pop $U$ from back\label{alg:removemax}
\WHILE{ $a_i > a_{\textrm{back}(U)}$}\label{alg:while1}
\STATE pop $U$ from back 
\ENDWHILE
\ELSE
\STATE pop $L$ from back \label{alg:removemin}
\WHILE{ $a_i < a_{\textrm{back}(L)}$} \label{alg:while2}
\STATE pop $L$ from back
\ENDWHILE
\ENDIF
\STATE append $i$ to $U$ and $L$\label{alg:append}
\IF{$i=w+\textrm{front}(U)$}  \label{alg:ensurescontainted}
\STATE pop $U$ from front
\ELSIF{$i=w+\textrm{front}(L)$}
\STATE pop $L$ from front
\ENDIF
\ENDFOR
 \end{algorithmic}
\end{small}
\caption{\label{algo:streaming}Streaming algorithm to compute the \textsc{max-min} filter using no more than 3~comparisons per element.
}
\end{algorithm}
%

%

\begin{proposition}\label{mainprop}
Algorithm~\ref{algo:streaming} computes the \textsc{max-min} filter over $n$ values using no more than $3n$~comparisons,
or 3~comparisons
per element.
\end{proposition}
\begin{proof}
We  prove by induction
that in Algorithm~\ref{algo:streaming}, $U$ and $L$ form a monotonic
wedge of $a$ over the interval $[\max\{i-w,1\},i)$ at the beginning of the main loop (line~\ref{alg:mainloop}).
Initially, when $i=2$, $U,L=\{1\}$, $U,L$ is trivially a monotonic wedge.
We have that the last component of both $U$ and $L$ is
$i-1$. If $a_i>a_{i-1}$ (line~\ref{alg:while1}), then we remove the last elements
of $U$ until $a_{\textrm{last}(U)}\geq a_{n+1}$ (line~\ref{alg:while1}) or if
$a_i\leq a_{i-1}$, we remove the last elements of 
$L$ until $a_{\textrm{last}(L)}\leq a_{n+1}$ (line~\ref{alg:while2}). Then we append
$i$ to both $U$ and $L$ (line~\ref{alg:append}). 
The lists $U,L$ form a monotonic 
wedge of $[\max\{i-w,1\},i]$ at this point (see \figref{mainalgo1}).
After appending the latest location $i$ (line~\ref{alg:append}),
any location $j<i$ will appear in either $U$ or $L$, 
but not in both. Indeed, $i-1$ is necessarily removed from either
$U$ or $L$. To compute the monotonic wedge over $[\max\{i-w+1,1\},i+1)$
from the monotonic wedge over $[\max\{i-w,1\},i]$,
we  check whether the location $i-w$ is in $U$ or $L$ at line~\ref{alg:ensurescontainted} and if so, we remove it.
Hence,
the algorithm produces the correct result.

\begin{figure}
\begin{center}
          \resizebox{0.8\columnwidth}{!} {
             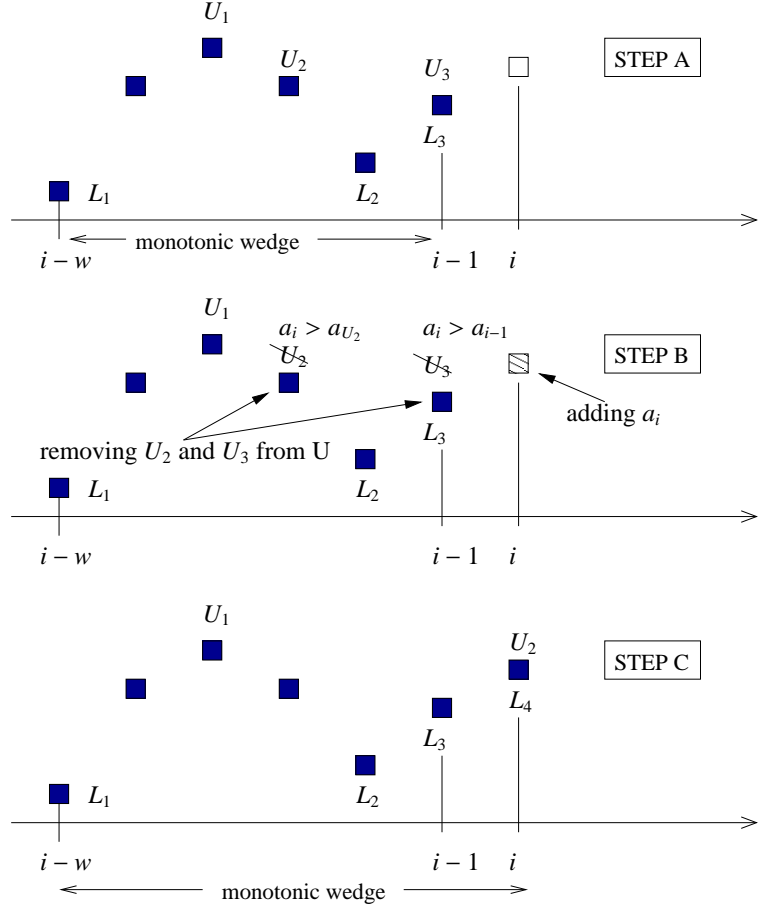
           }
\end{center}
\caption{
\label{mainalgo1}Algorithm~\ref{algo:streaming} from line~\ref{alg:mainloop}  to line~\ref{alg:append}: updating the monotonic wedge is done by either removing the last elements of $U$ or the last elements of $L$ until $U,L$ form a monotonic wedge for $[\max\{i-w,1\},i]$. }
\end{figure}



We still have  to prove that the algorithm will not use more than $3n$~comparisons, 
no matter what the input data is.
Firstly, the total number of elements that Algorithm~\ref{algo:streaming}
appends to queues $U$ and $L$ is $2n$, as each $i$ is appended both to $U$ and $L$ (line~\ref{alg:append}).
The comparison on line~\ref{alg:firstcompare} is executed $n-1$~time and each
execution removes an element from either $U$ or $L$ (lines \ref{alg:removemax} and
\ref{alg:removemin}), leaving $2n-(n-1)=n+1$~elements to be removed elsewhere.
 Because each time the comparisons on 
lines~\ref{alg:while1}  and \ref{alg:while2} gives {\em true}, an element is removed
from $U$ or $L$, there can only be  $n+1$ {\em true} comparisons. Morever,
the comparisons on   lines~\ref{alg:while1}  and \ref{alg:while2} can only be {\em false}
once for a fixed $a_i$ since it is the exit condition of the loop. The number
of {\em false} comparisons is therefore $n$. Hence, the total number of comparisons
is at most $(n-1)+(n+1)+n=3n$, as we claimed.
%
%
\end{proof}

While some signals such as electroencephalograms (EEG) resemble i.i.d noise, 
many more real-world signals are piecewise quasi-monotonic~\bracketcite{YLBICDMO05}.
While one \textsc{Gil-Kimmel} variant~\bracketcite{628856} has a comparison complexity of nearly
2~comparisons per element over i.i.d noise, but a worst case complexity of slightly more
than 3~comparisons for monotonic data, the opposite is true of our algorithm as demonstrated
by the following proposition.

\begin{proposition}
When the data is monotonic, Algorithm~\ref{algo:streaming} computes the \textsc{max-min} filter using no more than 2~comparisons per element.
\end{proposition}
\begin{proof}
If the input data is non-decreasing or non-increasing, then 
the conditions at line~\ref{alg:while1} and line~\ref{alg:while2} will never be true. Thus, in the worse case, for
each new element, there is one comparison at line~\ref{alg:firstcompare} and one at
either line~\ref{alg:while1} or line~\ref{alg:while2}.
\end{proof}

The next proposition shows that the memory usage of the monotonic
wedge is at most $w+1$~elements. Because $U$ and $L$ only store
the indexes, we say that
the total memory
buffer size of the algorithm is $2w+O(1)$~elements (see~\tableref{bigtable}).

\begin{proposition}
In Algorithm~\ref{algo:streaming}, the number of elements in the monotonic wedge
 ($\textrm{size}(U)+ \textrm{size}(L)$) is no more than $w+1$.
\end{proposition}
\begin{proof}
Each new element is added to both $U$ and $L$ at line~\ref{alg:append},
but in the next iteration of the main loop, this new element is
removed from either $U$ or $L$ (line~\ref{alg:removemax} or~\ref{alg:removemin}).
Hence, after line~\ref{alg:removemin} no element in the $w$ possible
elements can appear both in $U$ and $L$. Therefore
$\textrm{size}(U)+ \textrm{size}(L)\leq w+1$.
\end{proof}

\section{Implementation and Experimental Results}

While interesting theoretically, the number of comparison per element is not necessarily
a good indication of real-world performance. 
We implemented our algorithm in C++ using the STL \texttt{deque} template. 
A more
efficient data structure might be possible since the size of our 
double-ended queues are bounded by $w$. 
We used 64~bits floating point numbers (``double'' type).
In the pseudocode of Algorithm~\ref{algo:streaming}, we 
append $i$ to the two double-ended queues, and then we systematically pop one of them
(see proof of proposition~\ref{mainprop}).
We found it slightly faster to rewrite the code to avoid one pop and one append (see appendix).
The implementation of our algorithm stores only the location of the extrema whereas our
implementation of the \textsc{van Herk-Gil-Werman} algorithm stores values.
Storing locations means that we can compute the $\arg \max/\min$ filter with no overhead,
 but  each comparison is slightly more expensive. While our implementation
uses 32~bits integers to store locations, 64~bits integers should be
used when processing streams.
For small window sizes, \cite{628856} suggests unrolling
the loops, essentially compiling $w$ in the code: in this 
manner we could probably do away with a dynamic data structure
and the corresponding overhead.

We ran our tests on an AMD Athlon~64 3200+ using a 64~bit
  Linux platform with 1~Gigabyte of RAM 
(no thrashing observed).  The source code was
compiled using the GNU GCC 3.4 compiler with the optimizer option ``-O2''.

\begin{figure}
\begin{center}
\subfigure[\label{sinetimings}Input data is a sine wave with a period of 10~000 data points.]{
\includegraphics[width=0.65\columnwidth,angle=270]{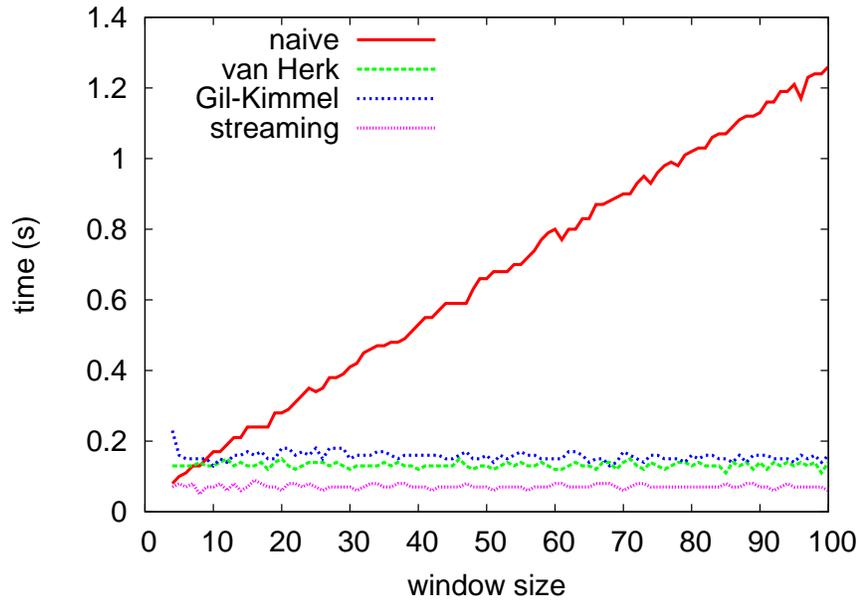}
}
\subfigure[\label{whitetimings}Input data is i.i.d. noise with a uniform distribution.]{
\includegraphics[width=0.65\columnwidth,angle=270]{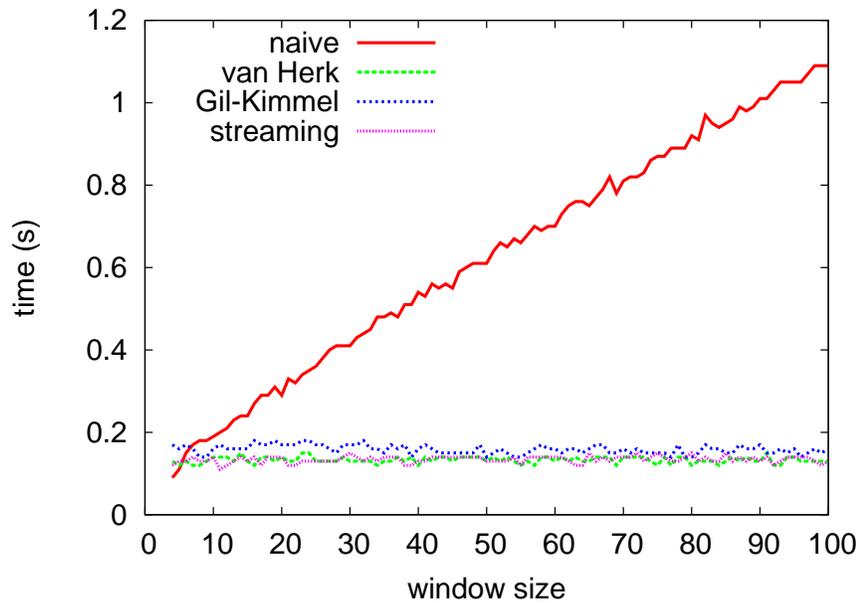}
}
\end{center}
\caption{
\label{timings}Running time to compute the \textsc{max-min} filter over a million data points using the naive  algorithm, our \textsc{van Herk-Gil-Werman} (van Herk) implementation, our \textsc{Gil-Kimmel} implementation, and our streaming implementation  (less is better). }
\end{figure}

We process synthetic data sets made of 1~million data points and report wall clock timings 
versus the window width (see \figref{timings}).
The linear time complexity of the naive algorithm is quite apparent  for $w>10$, but
for small window sizes ($w<10$), it remains a viable alternative.
Over i.i.d. noise generated with the Unix \texttt{rand} function, the \textsc{van Herk-Gil-Werman} and our algorithm are
comparable (see \figref{whitetimings}): both can process
1~million data points in about 0.15~s irrespective of the window width. 
For piecewise monotonic data such as a sine wave (see \figref{sinetimings})
our algorithm is roughly twice as fast and can process 1~million data points in about 0.075~s. 
Our C++ implementation of the \textsc{Gil-Kimmel} algorithm~\bracketcite{GilKimmelPatent} 
performed slightly worse than the \textsc{van Herk-Gil-Werman} algorithm.
To insure reproducibility, the source code is available freely from the author.


%

%
%

\section{Conclusion and Future Work}


We presented an algorithm to compute the \textsc{max-min} filter using no more
than 3~comparisons per element in the worst case whereas the previous best result was
slightly above $3 + 2 \log w /w +O(1/w)$~comparisons per element. 
Our algorithm has lower latency, is easy
to implement, and has reduced memory usage. For monotonic input, our algorithm
incurs a cost of no more than 2~comparisons per element.
Experimentally, our algorithm is especially competitive when the input is piecewise monotonic:
it is twice as fast  on a sine wave. 

We have shown that at least 2~comparisons per element are required to 
solve the \textsc{max-min} filter problem when no stream latency is allowed,
and we showed that this bound is tight when the window is small ($w=3$). 

%

\section*{Acknowledgements}

This work is supported by NSERC grant 261437. The author wishes to thank Owen Kaser
of the University of New Brunswick for his  insightful comments.

%

%


\bibliographystyle{njcarticle}
\bibliography{maxminalgo} 

%
\section*{Appendix: C++ source code for the streaming algorithm}
\lstset{language=C++, showstringspaces=false, breaklines,
 emph={U}, emphstyle={\color{blue}\textbf}, 
 emph={[2]L}, emphstyle={[2]\color{blue}\textbf}, 
emph={[3]maxval}, emphstyle={[3]\color{red}\textbf},
emph={[4]minval}, emphstyle={[4]\color{red}\textbf},
emph={[5]a}, emphstyle={[5]\textbf}} 

\begin{footnotesize}
\begin{lstlisting} 
// input: array a, integer window width w
// output: arrays maxval and minval
// buffer: lists U and L
// requires: STL for deque support
deque<int> U, L;
for(uint i = 1; i < a.size(); ++i) {
 if(i>=w) {
   maxval[i-w] = a[U.size()>0 ? U.front() : i-1];
   minval[i-w] = a[L.size()>0 ? L.front() : i-1];
 }// end if
 if(a[i] > a[i-1]) {
  L.push_back(i-1);
  if(i == w+L.front()) L.pop_front();
  while(U.size()>0) {
   if(a[i]<=a[U.back()]) {
    if (i == w+U.front()) U.pop_front();
    break;
   }// end if
   U.pop_back();
  }// end while
 } else {
  U.push_back(i-1);
  if (i == w+U.front()) U.pop_front();
  while(L.size()>0) {
   if(a[i]>=a[L.back()]) {
    if(i == w+L.front()) L.pop_front();
    break;
   }//end if
   L.pop_back();
  }//end while
 }// end if else  
}// end for
maxval[a.size()-w] = a[U.size()>0 ? U.front() : a.size()-1];
minval[a.size()-w] = a[L.size()>0 ? L.front() : a.size()-1];
\end{lstlisting}\end{footnotesize}
%

\end{document}